\documentclass[twoside]{article}
\usepackage{fleqn,espcrc2}
\input epsf
\input{epsf.sty}

\title{Magnetization Decay due to Vortex Phase Boundary Motion in BSCCO}

\author{M. Konczykowski$^{a}$, C. J. van der Beek$^{a}$, S.Colson$^{a}$,
M.V. Indenbom$^{a,b}$, P.H. Kes$^{c}$,Y. Platiel$^{d}$, E. 
Zeldov$^{d}$}

\address{Laboratoire des Solides Irradi\'{e}s, Ecole Polytechnique, 
91128 Palaiseau, France\\ 
\noindent $^{{\rm b}}$Institute of Solid State Physics, 142432
Chernogolovka, Moscow District, Russia\\
\noindent $^{{\rm c}}$Kamerlingh Onnes Laboratorium, Rijksuniversiteit Leiden, The Netherlands\\
\noindent $^{{\rm d}}$Department of Condensed Matter Physics, Weizmann 
 Institute of Science, Rehovot, Israel\\}

\begin{document}

\begin{abstract}
We identify a new regime of decay of the irreversible magnetization in clean Bi$_{2}$Sr$_{2}$CaCu$_{2}$O$_{8}$ 
crystals, at induction values close 
to the ``second peak field'' at which the bulk critical current density 
steeply increases. A time window is identified during which the decay of the  induction 
is controlled by the slow propagation of the phase transformation front across the sample.
\vspace{1pc}
\end{abstract}

\maketitle


The origin of the second magnetization peak (SMP) manifest in Bi$_{2}$Sr$_{2}$CaCu$_{2}$O$_{8}$ 
(BSCCO) crystals at low temperature is the object of major controversy.
Recent investigations of the role of weak disorder have confirmed its 
close relationship with the first order phase transition (FOT)
observed at higher temperatures. This has lead  to a generic phase diagram of vortex matter
\cite{Chikumoto92II,Khaykovich96,Khaykovich97,Vinokur98}, in which the 
SMP is thought to correspond to a phase transition from an ordered Bragg-glass at
 low vortex density (low fields) to a disordered vortex solid (glass) at 
 high fields. 
One of the key experiments in favor of a phase transition 
at the 

\begin{figure}[h]
    \vspace{-11pt}
    \centerline{\epsfxsize 6.7cm \epsfbox{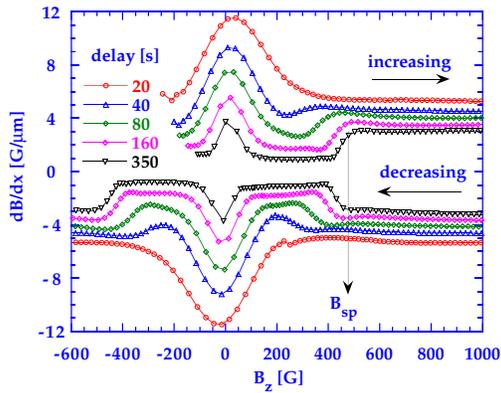}}
    \vspace{-18pt}
\caption{Hysteretic loops of the local induction gradient 
$\partial B / \partial x$ vs.  $B$, recorded at 15.9 K,
for various times after field application.}
\label{fig:Fig1}
\end{figure}

\newpage

\noindent SMP  was local magnetization, measured by the Hall-array 
technique \cite{Khaykovich96}. Such measurements show a sharp increase 
of the induction gradient $\partial 
B/\partial x$ at a well-defined value of the induction $B_{sp}$, corresponding to 
the value at which the transition takes place. 
Here, we show from time resolved measurements of the 
induction 
that there exists a time regime during which the slow motion of the phase transformation front determines the global
magnetic relaxation.


The lightly oxygen overdoped BSCCO crystal used in this study, cut from a 
larger crystal, was checked for its uniformity by magneto-optical imaging 
of the flux penetration.  Local induction profiles were measured on the crystal 
surface using a 2D electron gas Hall--probe array, composed of 11 sensors 
of area $10 \times 10$ $\mu$m$^{2}$, and spaced by 10 $\mu$m.
Increasing (decreasing) branches of hysteretic ``local magnetization'' 
loops at various waiting times were obtained by swiftly ramping the 
applied field $H_{a}$ up (down)  to its target value from a starting point
lower (higher) by several times the full penetration field, after which the magnetic relaxation was
recorded during 350 s. On the experimental time scale  ($t > 5$ s), the SMP at 
$T \raisebox{1mm}{$>$}\!\!\!\!\!\raisebox{-1mm}{$\sim$} 25$ K is marked by the 
emergence of Bean-like induction profiles from 
dome-shaped profiles related to screening by the surface barrier
currents. Only at temperatures below 20 K can one observe the SMP as
a crossover from one bulk pinning regime to another. Fig.~1 shows loops 
of

\begin{figure}[t]
\centerline{\epsfxsize 6.7cm \epsfbox{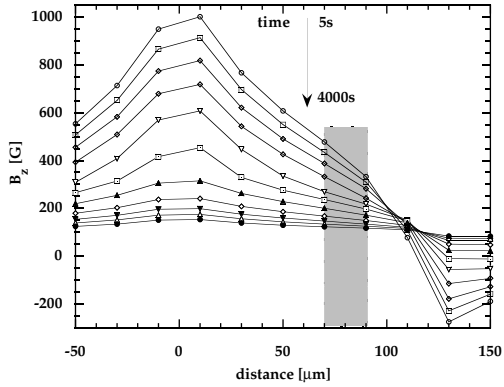}}
\vspace{-14pt}

\caption{Decay of the magnetic induction on the surface of the BSCCO sample
at $T = 15.9$ K and $H_{a} = 100$ Oe, after field-cooling in 3 kOe followed 
by the fast decrease of the applied field.}
\label{fig:Fig2}
\vspace{-20pt}
\end{figure}

\noindent   $\partial B/\partial x$ vs. 
the local induction $B$ at various times after the end of the field
ramp, for $T = 15.9$ K. The characteristic jump of $\partial B / 
\partial x$ at $B_{sp}$ appears only at long times; at short times,
{\em i.e.} at higher screening current, the inverse situation occurs and $\partial B/\partial x$
is higher for $ B  <  B_{sp}$. 

Long time relaxations at various applied fields $H_{a}$ were recorded
by cooling the sample down from $ T > 30$ K in a field
exceeding $H_{a}$ by 3 kOe. After reaching thermal stability at 15.9 K, the field was
rapidly decreased to $H_{a}$ and the decay of the magnetic induction profiles
was recorded during 4000 s (Fig.~2).
From the spatially resolved magnetic relaxation, the electric field is obtained by summing 
the numerically calculated time derivatives $\partial B / \partial t$ 
for five sensors, from the center of the crystal outwards. 
Assuming that $\partial B /\partial x$ is proportional to the local current 
density $j$, and that the electric field 
arises as a result of thermally activated
vortex creep Ref.~\cite{Abulafia96}, a plot of 
$U\propto{kT\ln(E/B/j)}$ against $\partial B / \partial x$  represents
the variation of the flux creep energy barrier with current density  (Fig. 3).  
When $H_{a} > B_{sp}$, ($H_{a}\geq 500 Oe$), smooth, power like variations 
$U \propto j^{-0.38}$ are obtained. A completely new behavior is observed when 
$H_{a} < B_{sp}$: then, the divergence of $U(j)$ stops at a given 
time, after which an extended plateau in the  $U(j)$ curve appears. 
The correlation of the $U(j)$--curves with

\begin{figure}[t]
    \centerline{\epsfxsize 6.7cm \epsfbox{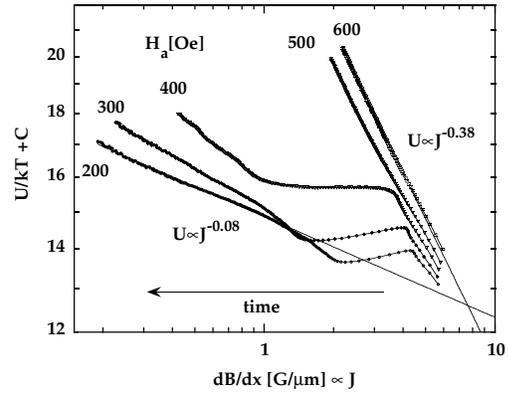}}
\vspace{-18pt}
\caption{Flux creep activation barrier vs. current variations $U(j)$ 
in the BSCCO crystal at 15.9 K, for various applied fields.}
\label{fig:Fig3}
\vspace{-22pt}
\end{figure}

\noindent   the decaying flux profiles shows 
that the beginning of the plateau at high $j$ corresponds to the 
first appearance of the low--field ordered vortex phase (identifiable 
in Fig.~2 by the smaller gradient $\partial B/ \partial x$) in the crystal. 
The end of the plateau at long $t$ or small $j$ occurs when there is 
no region of high--field phase (corresponding to the region of higher $\partial B/ \partial x$)
left in the sample. Only then does a new power-law--like divergence of 
$U \sim j^{-0.08}$, representative of the activation barriers in the 
low--field phase, start. In the intermediate time interval,
the relaxation process is determined by the slow motion of phase transformation front across
the sample.


Concluding, the low--field and high--field vortex phases 
are characterized by very different $U(j)$-- (or $I(V)$--) relations. 
The SMP occurs as a result of the jump from one  $I(V)$--curve to another. In the regime of phase 
coexistence at $B \sim B_{sp}$, the electrodynamics of the sample is 
determined by the motion of the phase transformation front. 


\end{document}